\documentclass[10pt,peerreview,draftcls,onecolumn]{IEEEtran}

\usepackage{graphicx}
\usepackage{url}
\usepackage{epstopdf}
\usepackage{color}
\usepackage{xcolor}
\usepackage{bbm}
\usepackage[cmex10]{amsmath}
\usepackage{array}
\usepackage{caption}
\colorlet{mygreen}{green!60!gray}

\begin{document}
%
\title{Optimum Fusion of Possibly Corrupted Reports for Distributed Detection in Multi-Sensor Networks}
%
%
%

\author{Andrea Abrardo,~ Mauro Barni,~Kassem Kallas,~Benedetta Tondi
%

}
\maketitle

\begin{abstract}
The most common approach to mitigate the impact that the presence of malicious nodes has on the accuracy of decision fusion schemes consists in  observing the behavior of the nodes over a time interval $T$ and then removing the reports of suspect nodes from the fusion process. By assuming that some a-priori information about the presence of malicious nodes and their behavior is available, we show that the information stemming from the suspect nodes can be exploited to further improve the decision fusion accuracy. Specifically, we derive the optimum fusion rule and analyze the achievable performance for two specific cases. In the first case, the states of the nodes (corrupted or honest) are independent of each other and the fusion center knows only the probability that a node is malicious. In the second case, the exact number of corrupted nodes is fixed and known to the fusion center. We also investigate the optimum corruption strategy for the malicious nodes, showing that always reverting the local decision does not necessarily maximize the loss of performance at the fusion center.
\end{abstract}


\IEEEpeerreviewmaketitle

\section{Introduction}
\label{sec.intro}

Distributed detection in the presence of corrupted nodes, often referred to as byzantine nodes \cite{Vemp13}, or simply Byzantines, has received an increasing attention for its importance in several application scenarios, including wireless sensor networks \cite{WSNvarshney,WSNanomalyDet}, cognitive radio \cite{WLSH10,Raw11}, distributed detection \cite{Mar09}, multimedia forensics \cite{Bar13} and many others.

The most commonly studied scenario is the parallel distributed data fusion model. According to such a model, the $n$ nodes of a multi-sensor network observe a system over a period of time $T$ through the vectors $x_1(t), x_2(t) \dots x_n(t)$, $t = 1 \dots T$. Based on such vectors, the nodes compute $n$ reports and send them to a Fusion Center (FC). The fusion center gathers the local reports and makes a final decision about the status of the observed system at each time instant $t$. Hereafter, we assume that the system can be only in two states ($0$ and $1$), additionally we make the simplifying assumption that the reports correspond to binary values whereby the local nodes inform the fusion center about their knowledge on system status. Specifically, we indicate by $r_{ij} \in \{0,1\}, i = 1 \dots n, j = 1 \dots T$ the report sent by node $i$ at time $j$.

\begin{figure}[t!]
\centering
    \includegraphics[width=0.4\textwidth]{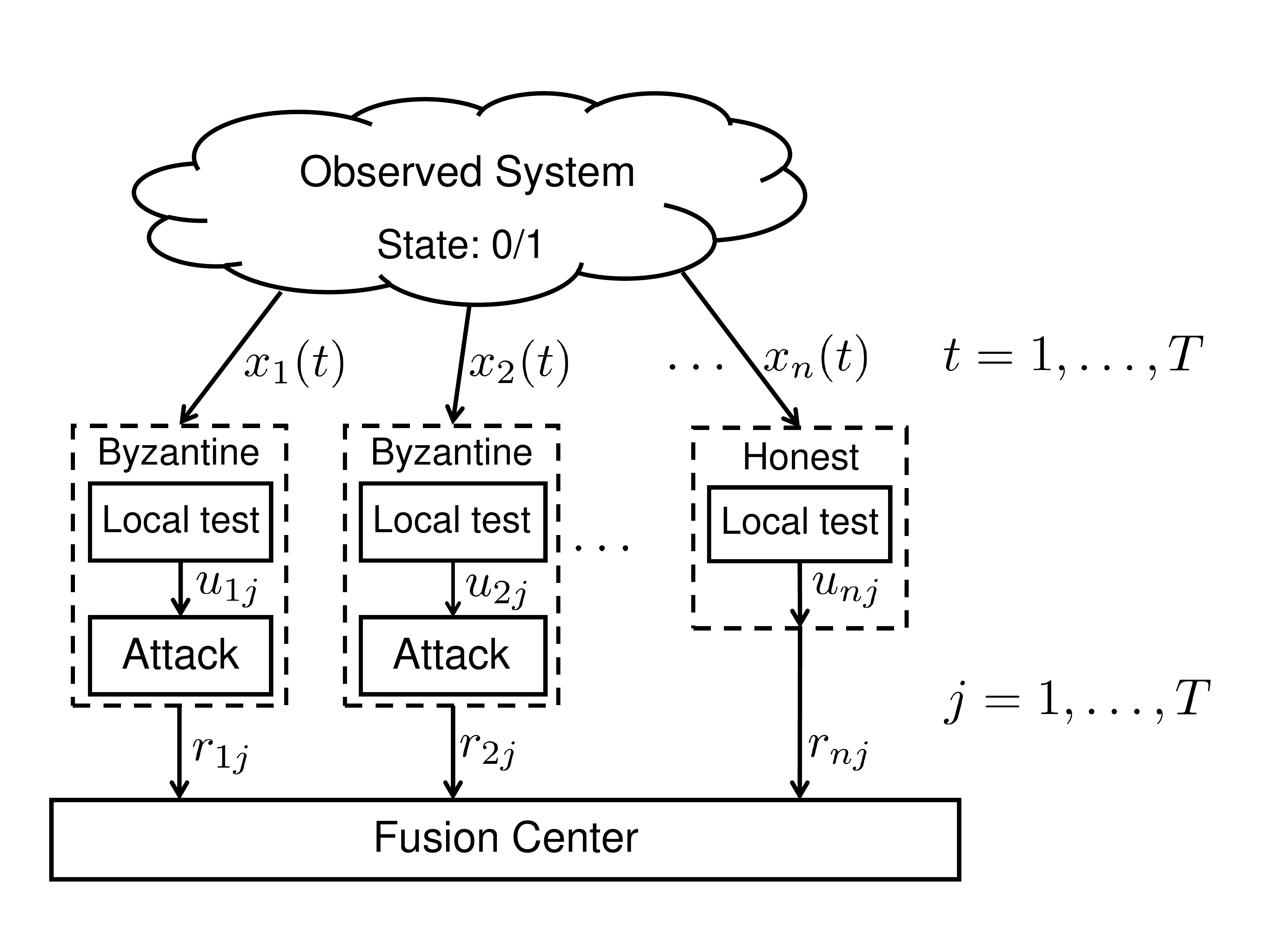}
    \caption{Sketch of the adversarial decision fusion scheme. \vspace{-0.5cm}}
    \label{fig.setup}
\end{figure}

Decision fusion must be carried out in an adversarial setting, that is by taking into account the possibility that some of the nodes malevolently alter their reports to induce a decision error. This is a recurrent situation in many scenarios wherein the nodes may make a profit from a decision error (see \cite{Vemp13} for a general introduction to this topic). More rigorously, since the nodes do not know the exact system status they must estimate it based on the observation vectors $x_i(t)$. Let us denote by $u_{ij}$ the binary local decision made by node $i$ regarding the status of the system at time $j$. While for honest nodes $r_{ij} = u_{ij}$, we assume that malicious nodes flip $u_{ij}$ with a certain probability $P_{mal}$. In such a case we obviously have $u_{ij} \ne r_{ij}$. Hereafter, we assume that the state of each node (honest or byzantine) remains constant over the whole observation period $T$. In addition, we make the simplifying assumption that the flipping probability does not depend on the observed sequence $x_i(t)$.
A pictorial representation of the setup described above is given in figure \ref{fig.setup}.
If we assume that the FC makes the decision on system status at time $t$ by considering only the reports corresponding to $j = t$, then the optimum way to combine the local decisions according to the Bayesian approach has already been determined in \cite{OptFusion,Var97}, and is known as Chair-Varshney rule. In the case of symmetric error probabilities at the nodes, Chair-Varshney rule corresponds to simple majority-bases decision.
Better results can be obtained by assuming that the FC collects the reports received from the nodes over a time window T, and decides on all the states assumed by the system at the end of the observation interval. Given that node states are constant over $T$, it is easy to argue that observing the reports from the nodes over many instants allows the FC to identify byzantine nodes and hence reduce the error probability of the final decision.

\section{Prior work and contribution}

The problem described in the previous section, along with several variants, has already been studied in several papers, including \cite{Mar09, Raw11,CDC, Bar13}. In the great majority of the cases, the fusion center observes the node reports over the time window $T$ trying to identify byzantine nodes. Once the byzantine nodes have been identified, the fusion center {\em isolates} them by removing their reports from the fusion process.
%
The most relevant works for this paper are \cite{Raw11,CDC}, which generalize the analysis carried out in \cite{Mar09}. Even in these works the malicious nodes do not take advantage of the knowledge of the observed sequences $x_i(t)$, do not cooperate with each other, and do not know the true state of the system. In contrast to \cite{Mar09}, the analysis is not limited to the asymptotic case. The papers propose two methods to identify the byzantine nodes by observing the behavior of the nodes over the time window $T$ and assigning them a reputation measure. Then, decision fusion is carried out by discarding the nodes with a low reputation.  A different isolation scheme based on adaptive learning is described in \cite{LearnByzantines}, where the observed behavior of the nodes over time is compared with the expected behavior of honest nodes. A peculiarity of this scheme is that it works even when the majority of the nodes are byzantine, but it requires very large T to achieve good performances. Tolerant schemes which mitigate the impact of Byzantines in the decision have also been proposed, like in \cite{tolerant_scheme}, where the reports are weighted differently according to node reliability but no one is removed from the network.

In this paper, we start from the observation that in the presence of some a-priori information about the probability distribution of Byzantine nodes and their behaviour, removing the reports received from suspect nodes is not the optimum strategy, since such reports may still convey some useful information about the status of the observed system. Specifically, we derive the optimum fusion rule, by assuming that the probability distribution of malicious nodes across the network and the flipping probability $P_{mal}$ are known. We analyze the performance of the system for two particular setups: according to the first one the states of the nodes (corrupted or honest) are independent of each other and the fusion center knows only the probability that a node is malicious. In the second case, the exact number of corrupted nodes is known to the fusion center.  Having derived the optimum fusion strategy, we turn our attention to the optimum attacking strategy for the Byzantines, in term of flipping probability, showing that in some cases always flipping the local decision ($P_{mal} = 1$) is not the best choice.

The rest of this paper is organized in two main sections: in Section \ref{sec.PRstates}, we derive the optimum fusion rule, while in Section \ref{sec.simul} we use simulations to analyze the performance of the optimum fusion rule.
\section{Optimum fusion rule}
\label{sec.PRstates}

In the rest of the paper, we will use capital letters to denote random variables and lowercase letters for their instantiations.
Let $S^T = (S_1, S_2 \dots S_T)$, with $S_i \in \{0,1\}$, be a sequence of independent and identically distributed (i.i.d.) random variables defining the states assumed by the system. We assume all the states to be equiprobable.
We indicate by $U_{ij} \in \{0,1\}$ the local decision made by node $i$ about the system status at time $j$. We exclude any interaction between the nodes and assume that $U_{ij}$'s are conditionally independent for a fixed status of the system. This is equivalent to assuming that the local decision errors are i.i.d.

It is assumed that the state of the nodes in the network is not known to the FC, which however has some prior statistical knowledge about the presence of Byzantines. To elaborate, let $A^n = (A_1 \dots A_n)$ be a binary random vector describing the states of the nodes over the network. More specifically, each random value $A_i$ denotes the state of node $i$: $A_i = 0$ (res. $A_i = 1$) indicates that node $i$ is honest (res. byzantine). In the scenario considered in this paper, the FC is supposed to know the prior probability  of $A^n$, i.e. $P_{A^n}(a^n)$ or simply $P(a^n)$.

Finally, we let ${\bf R} = \{R_{ij}\}, ~ i = 1 \dots n, j = 1 \dots T$ be a random matrix with all the reports received by the fusion center, and ${\bf r} = r_{ij}$ a specific instantiation of such a matrix. As stated in the introduction, $R_{ij} = U_{ij}$ for honest nodes, while $P(R_{ij} \ne U_{ij}) = P_{mal}$ for byzantine nodes. According to our model, local decisions $U_{ij}$ are flipped independently of each other with equal probabilities, so that the action of malicious nodes can be modeled as a number of independent binary symmetric channels with crossover probability $P_{mal}$.

We are now ready to derive the optimum decision rule at the FC. Given the received reports ${\bf r}$ and by adopting a maximum a posteriori probability criterion, the optimum decision rule which minimizes the error probability $P_e$ can be written as:

\begin{equation}
s^{T,*} = \arg\max_{s^T} P(s^T | {\bf r}).
\label{eq.map}
\end{equation}

By applying Bayes rule and exploiting the fact that the all state sequences are equiprobable we get:
%
\begin{equation}
s^{T,*} =  \arg\max_{s^T} P({\bf r} | s^T ).
\label{eq.ML}
\end{equation}
Since in our model the states of the nodes are themselves random variables, we can write:

\begin{align}
s^{T,*}  
= & \arg\max_{s^T} \sum_{a^n} P({\bf r} | a^n, s^T) P(a^n) \label{eq.pseudoML}\\
= & \arg\max_{s^T} \sum_{a^n} \bigg(\prod_{i=1}^n P(r_i | a_i, s^T )\bigg) P(a^n)\label{eq.pseudoML_2}\\
= & \arg\max_{s^T} \sum_{a^n} \bigg(\prod_{i=1}^n \prod_{j=1}^T P(r_{ij} | a_i, s_j )\bigg) P(a^n),\label{eq.pseudoML_3}
\end{align}

where $r_i$ indicates the $i$-th row of ${\bf r}$. In \eqref{eq.pseudoML_2} we exploited the fact that, given the state of the nodes and the states of the system over $T$, the reports sent by the nodes are independent of each other, while \eqref{eq.pseudoML_3} derives from the observation that, given the node state, each report depends only on the corresponding status of the system.

We now consider the case in which the probability of a local decision error, say $\varepsilon$, is the same regardless of the system status, that is $\varepsilon = P(U_{ij} \neq S_j|S_j = s_j)$, $s_j = 0,1$. For a honest node such a probability corresponds to the probability that the report received by the FC does not correspond to the system status. This is not the case for byzantine nodes, for which the probability of a wrong report, hereafter indicated by $\delta$, can be written as $\delta = \varepsilon (1 - P_{mal}) + (1 - \varepsilon)P_{mal}$.

According to the above setting, the nodes can be modeled as binary symmetric channels, whose input corresponds to the system status and for which the crossover probability is equal to $\varepsilon$ for the honest nodes and $\delta$ for the byzantine nodes. With regard to $\varepsilon$ we can safely assume that such a value is known to the fusion center. The value of $\delta$ depends on the value of  $P_{mal}$ which is chosen by the Byzantines and then should be optimally determined in a game-theoretic framework (see \cite{KBKV13} for an example in this sense). While this is an interesting possibility, we leave such an analysis to a future work, and assume that $P_{mal}$ is known to the FC.
From \eqref{eq.pseudoML_3}, the optimum decision rule becomes:
\begin{align}
\label{eq.pseudoML_symm}
s^{T,*} =  \arg\max_{s^T} & \sum_{a^n} \bigg(\prod_{i:a_i = 0}  (1-\varepsilon)^{n_{eq}(i)} \varepsilon^{T-n_{eq}(i)} \bigg. \\ & \hspace{0.5cm} \bigg. \prod_{i:a_i = 1} (1-\delta)^{n_{eq}(i)} \delta^{T-n_{eq}(i)} \bigg) P(a^n),\nonumber
\end{align}
where $n_{eq}(i)$ is the number of $j$'s for which $r_{ij} = s_j$.

In Section \ref{sec.OF_Random} and \ref{sec.OF_DETstates} we consider two simple, yet insightful, cases of prior knowledge on the distribution of the states of the nodes in the network.

\subsection{Independent node states}
\label{sec.OF_Random}

In this section we consider a situation in which the states of the nodes are independent of each other and the state of each node is described by a Bernoulli random variable with parameter $\alpha$, that is $P(A_i = 1) = \alpha$, $\forall i$. In this way, the number of byzantine nodes in the network is a random variable with a binomial distribution.
%
%
%
%
%
In particular, we have $P(a^n) = \prod_i P(a_i)$, and hence \eqref{eq.pseudoML_2} can be rewritten as:
\begin{equation}
s^{T,*} = \arg\max_{s^T} \sum_{a^n} \bigg(\prod_{i=1}^n P(r_i | a_i, s^T )P(a_i) \bigg).
\label{eq.factorization}
\end{equation}
The expression in round brackets corresponds to a factorization of $P({\bf r}, a^n | s^T )$.
If we look at that expression as a function of $a^n$, it is a product of marginal functions.
By exploiting the distributivity of the product with respect to the sum
we can rewrite \eqref{eq.factorization} as follows
\begin{equation}
s^{T,*} =  \arg\max_{s^T} \prod_{i=1}^{n} \bigg(\sum_{a_i} P( r_i| a_i, s^T) P(a_i) \bigg),
\label{eq.pseudoML_Random}
\end{equation}
which can be computed more efficiently, especially for large $n$.
The expression in equation \eqref{eq.pseudoML_Random} can also be derived directly from \eqref{eq.ML} by exploiting first the independence of the reports and later applying the total probability law.
Since, as noticed before, $P( r_i| a_i, s^T) = \prod_{j=1}^T P( r_{ij}| a_i, s_j)$, from \eqref{eq.pseudoML_Random}
it is easy to derive the to-be-maximized expression for the case of symmetric error probabilities at the nodes, which is
\begin{align}
s^{T,*} = & \arg\max_{s^T} \prod_{i=1}^{n} \left[(1-\alpha)(1-\varepsilon)^{n_{eq}(i)} \varepsilon^{T-n_{eq}(i)}\right.\nonumber\\
 & \hspace{3cm} \left.+ \alpha(1-\delta)^{n_{eq}(i)} \delta^{T-n_{eq}(i)}\right].
 \label{eq.pseudoML_Random_2}
\end{align}

%
%

\subsection{Fixed number of Byzantines}
\label{sec.OF_DETstates}

In this section we consider a scenario in which the exact number of Byzantines in the network is known to the FC. Let us indicate such a number with $k$ (to draw a parallelism with the analysis carried out in the previous sections, $k$ should be equal to the average number of Byzantines nodes that is $k = \alpha n$). Since the overall number of Byzantines is fixed, the state of a node is no longer independent from the states of the other nodes. More specifically, we have $P(a^n) = {\binom{n}{k}}^{-1}$ if the numbers of 1's in  $a^n$ is $k$, 0 otherwise, since the only configurations with non-zero probability are those with exactly $k$ Byzantines.
Let $\mathcal{I}$ be the indexing set $\{1,2,...,n\}$.  We denote with $\mathcal{I}_k$ the set of all the possible $k$-subsets of $\mathcal{I}$. Let $I \in \mathcal{I}_k$ be a random variable with the index of the Byzantine nodes, a node $i$ being byzantine if $i \in I$, honest otherwise.
Then, from \eqref{eq.pseudoML} we derive

\begin{align}
s^{T,*} = \arg\max_{s^T}  \sum_{I \in \mathcal{I}_{k}} P({\bf r} | I, s^T ) p(s^T),\label{eq.ML_Determ}
\end{align}
which in the symmetric case corresponds to
\begin{align}
s^{T,*} =  \arg\max_{s^T} & \sum_{I \in \mathcal{I}_{k}} \bigg(\prod_{i \in I} (1-\delta)^{n_{eq}(i)} \delta^{T-n_{eq}(i)}\bigg.\nonumber\\
  & \hspace{1cm} \bigg.\prod_{i \in \mathcal{I} \setminus I} (1-\varepsilon)^{n_{eq}(i)} \varepsilon^{T-n_{eq}(i)}\bigg).
  \label{eq.ML_Determ_2}
\end{align}
We conclude this section observing that implementing equation \eqref{eq.ML_Determ_2} is more complex than \eqref{eq.pseudoML_Random_2}, due to the summation over all the possible $k$-subsets of $\mathcal{I}$.

\begin{table*}[t!]
\small
\centering
\renewcommand{\arraystretch}{1.1}
\begin{tabular}{c| c c c c c c|}
                       \hline
\multicolumn{1}{|c|}{Setting/$P_{mal}$} & 0.5   & 0.6   & 0.7   & 0.8   &0.9   &1.0   \\ \hline
\multicolumn{1}{|l|}{$n=16$, $\alpha=0.4375$, $T=4$}  &0.0131 &0.0221 &0.0374	&0.0777	&0.1853 &{\bf 0.3162}\\ \hline
\multicolumn{1}{|l|}{$n=11$, $\alpha=0.4545$, $T=9$} &0.0217 &0.0278 &0.0302	&0.0444	&0.1320 &{\bf 0.3708}\\ \hline
\multicolumn{1}{|l|}{$n=10$, $\alpha=0.4$, $T=10$}  & 0.0176 &0.0211 &0.0200 &0.0311 &0.1003 &{\bf 0.2663}\\ \hline
\multicolumn{1}{|l|}{$n=5$, $\alpha=0.4$, $T=15$} &0.0814 &0.0951 &0.0919 &0.0869 &0.1640 &{\bf 0.3189}\\\hline
\end{tabular}

\caption{$P_{e,R}$ versus $P_{mal}$ under different settings.}
\label{tab.ProbTable}
\end{table*}

\begin{table*}[t!]
\small
\centering
\renewcommand{\arraystretch}{1.1}

\begin{tabular}{c| c c c c c c|}
  \hline
\multicolumn{1}{|c|}{Setting/$P_{mal}$} &0.5   &0.6   &0.7   &0.8   &0.9   &1.0   \\ \hline
\multicolumn{1}{|l|}{$n=16$, $\alpha=0.4375$, $T=4$}  &0.0045  &0.0054  &0.0042  &0.0067  &0.0190  &{\bf 0.0357}\\ \hline
\multicolumn{1}{|l|}{$n=11$, $\alpha=0.4545$, $T=9$} &{\bf 0.0093} &0.0090  &0.0058  &0.0043  &0.0048  &0.0046\\ \hline
\multicolumn{1}{|l|}{$n=10$, $\alpha=0.4$, $T=10$}  &{\bf 0.0101}  &0.0079  &0.0060  &0.0038  &0.0023  &0.0011\\ \hline
\multicolumn{1}{|l|}{$n=5$, $\alpha=0.4$, $T=15$} &{\bf 0.0339}  &0.0301  &0.0297  &0.0294  &0.0177  &0.0087\\ \hline
\end{tabular}
\caption{$P_{e,F}$ versus $P_{mal}$ under different settings.}
\label{tab.DetTable}
\end{table*}

\section{Simulation results and discussion}
\label{sec.simul}

We run simulations to estimate the error probability at the FC for both the case of independent node states and fixed number of nodes. Let $P_{e,R}$ and $P_{e,F}$ be the error probabilities for the former and latter case respectively.
The simulations were carried by considering networks with different number of nodes and various values of $T$ in order to test the effect of these parameters on the detection performance as well as on the optimum strategy for the malicious nodes.
\begin{figure}[t!]
\centering
\includegraphics[width=0.45\textwidth,height=4.5cm]{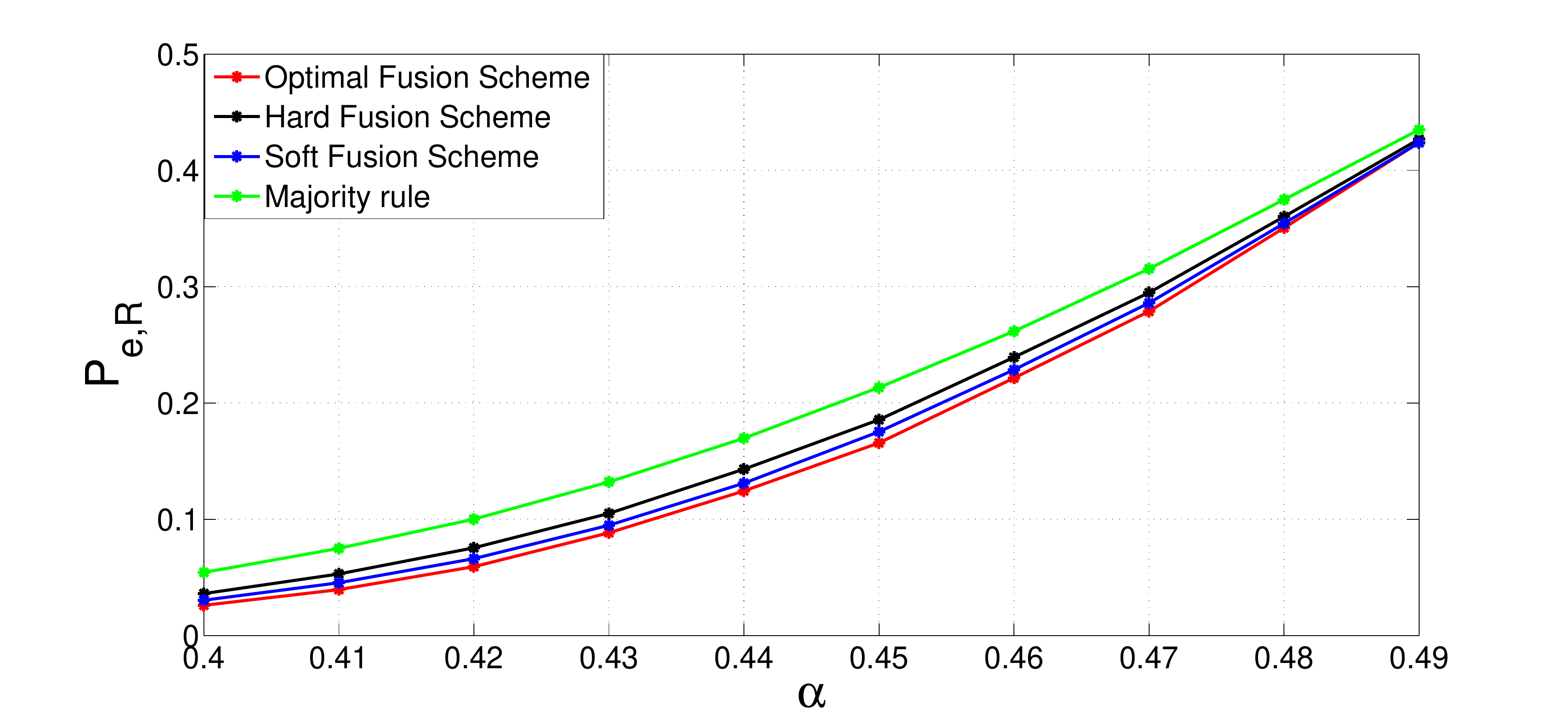}
    \caption{$P_{e,R}$ vs $\alpha$ ($P_{mal}=1$, $\varepsilon = 0.1$, $n=100$, $T=4$) \vspace{-0.7cm}}.
    \label{fig.perralfa}
\end{figure}
\begin{figure}[t!]
\centering
\includegraphics[width=0.45\textwidth,height=4.5cm]{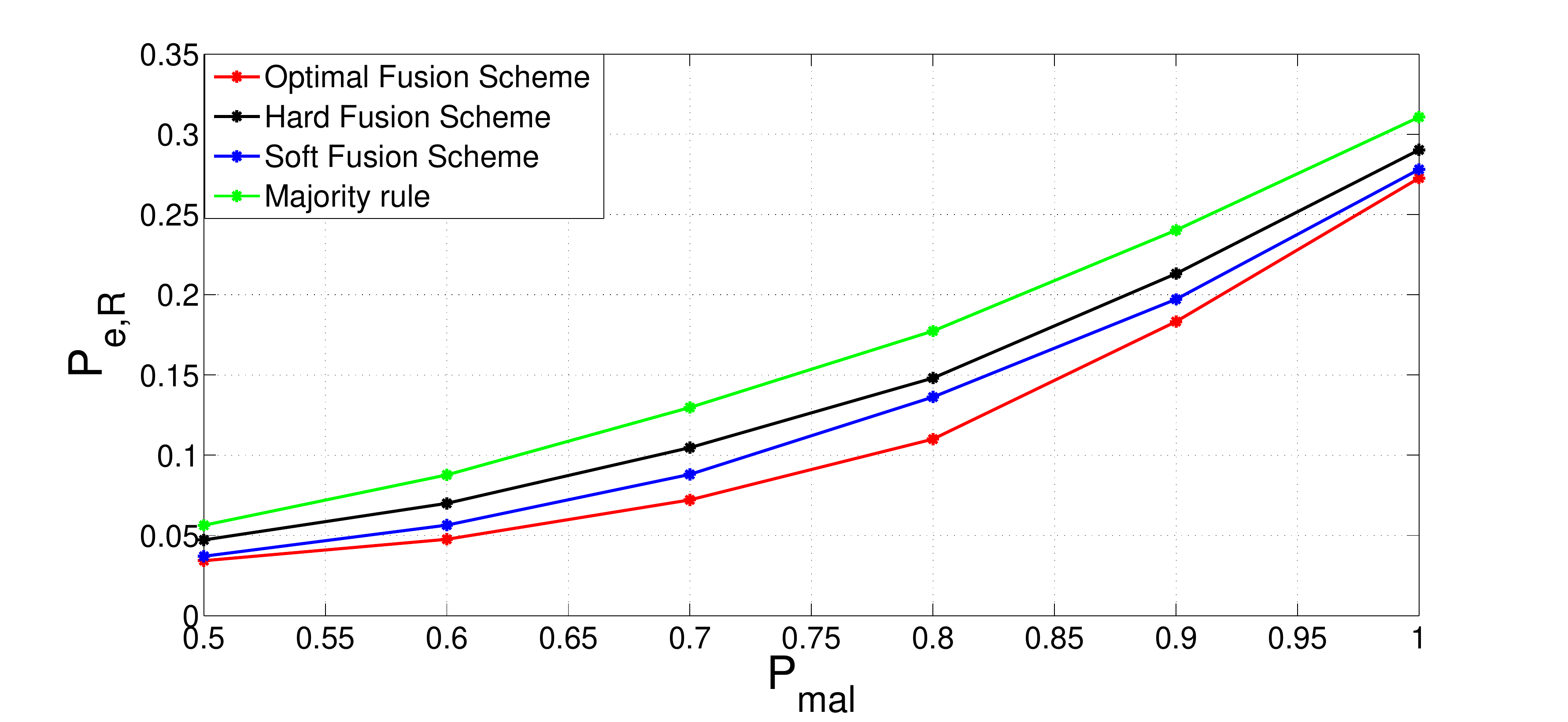}
    \caption{$P_{e,R}$ vs $P_{mal}$ ($\alpha=0.4$, $\varepsilon = 0.1$, $n=10$, $T = 4$).\vspace{-0.5cm}}
    \label{fig.perrpmalN10}
\end{figure}
\begin{figure}[t!]
\centering
\includegraphics[width=0.45\textwidth,height=4.5cm]{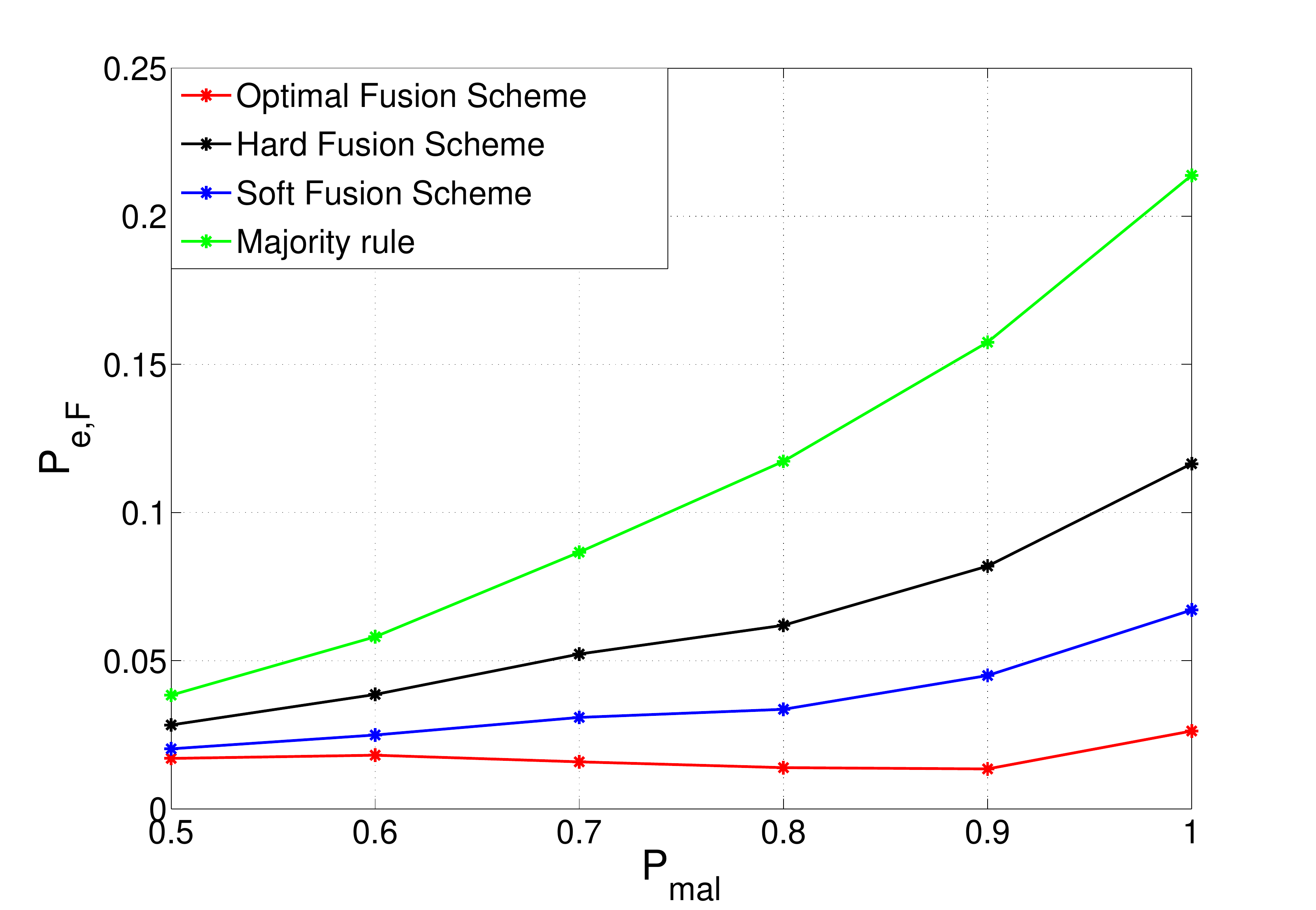}
    \caption{$P_{e,F}$ vs $P_{mal}$ ($\alpha=0.4$, $\varepsilon = 0.1$, $n=10$, $T = 4$).\vspace{-0.5cm}}
    \label{fig.perrDpmal}
\end{figure}

In Figure \ref{fig.perralfa}, we simulated a network with $n=100$ nodes, with independent states. The simulations were carried out by letting $P_{mal} = 1$ and $\alpha$ ranging in the interval $[0.4, 0.49]$. Upon inspection of the results, we can see that the optimal fusion rule achieves a lower $P_{e}$ compared to the majority rule and the isolation schemes proposed in \cite{Raw11} and \cite{CDC}.

In Figure \ref{fig.perrpmalN10}, we simulated a smaller network with $n=10$ nodes, fixing $\alpha=0.4$ and letting $P_{mal}$ vary in $[0.5,1.0]$. With such a small number of nodes, we were able to also simulate a network with a fixed number of nodes ($k=4 = \alpha n$), as shown in   Figure \ref{fig.perrDpmal}. As a first observation, in both cases the optimum fusion rule outperforms the other rules achieving lower $P_e$. Moreover, the benefit from the optimum fusion rule is more evident when the number of Byzantines is fixed (in fact in this case the FC has more information about the distribution of malicious nodes over the network).
With regard to $P_{mal}$, we can see that for the setting used in the figures, the optimum attacking strategy for the malicious nodes corresponds to always flipping the result of the local decision. Such a result agrees with previous works in \cite{Raw11,CDC}.

As a further analysis, we investigated the impact of $T$ on the error probability. The results we obtained are given in Table \ref{tab.ProbTable} and Table \ref{tab.DetTable}. Due to complexity of the optimum decision fusion rule for the case of a fixed number of nodes, we had to limit the size of the simulations by letting $n+T=20$.
By looking at Table \ref{tab.ProbTable}, we see that, in the case of independent node states, the preferable strategy for the Byzantines is always $P_{mal}=1.0$. Once again this agrees with  the analysis carried out in \cite{CDC},\cite{KBKV13}.
This is no more the case when the number of nodes is fixed. Table \ref{tab.DetTable} shows that when the observation windows $T$ increases ($T = 9, 10, 15$ in the table), $P_{mal}=1$ is no longer the optimal attack, which now corresponds to $P_{mal} = 0.5$. The intuition behind this apparently unexpected behavior is that when the byzantine nodes can be identified, and since the FC is assumed to know $P_{mal}$, the FC can revert the action of the Byzantines by simply inverting the reports received from such nodes. This is what the optimal fusion rule implicitly does. Interestingly such a behavior becomes evident in the case of fixed number of Byzantines only (and for a large enough observation window), since in such a case the a-priori information available to the FC is greater hence easing the identification of the malicious nodes. This also explains why in this case the optimum strategy for the Byzantines corresponds to $P_{mal} = 0.5$. With such a strategy, in fact, the mutual information between the system status and the reports is zero, thus minimizing the information the FC can rely on. Even if the above results refer only to a limited number of cases, they lead us to conjecture the existence of a bimodal behavior for the optimum corruption strategy: when the FC can not identify reliably the byzantine nodes, try to directly induce a decision error by letting $P_{mal} = 1$, otherwise it simply minimize the information available to the FC by letting $P_{mal} = 0.5$.

\section{Conclusions}
\label{sec.conc}

We derived the optimum fusion rule for a FC receiving possibly corrupted reports regarding the status of an observed system. We considered two different setups corresponding to different a-priori information about the distribution of byzantine nodes. We showed the advantage of the optimum fusion strategy with respect to previous works through simulations, and identified a bimodal behavior of the optimum flipping probability to be adopted by the Byzantines. In future works we will try to remove some of the hypotheses behind our analysis, namely the necessity that the FC knows $P_{mal}$ and the assumption that the byzantine nodes do not exploit the knowledge of $x_i(t)$ to devise a more powerful attack. Casting the fusion problem into a game-theoretic setting where the FC and the Byzantines pursue opposite goals along the lines highlighted in \cite{BarGon13} is also an interesting research direction.
\bibliographystyle{IEEEtran}
\bibliography{SPLbiblio}

\begin{thebibliography}{10}
\providecommand{\url}[1]{#1}
\csname url@samestyle\endcsname
\providecommand{\newblock}{\relax}
\providecommand{\bibinfo}[2]{#2}
\providecommand{\BIBentrySTDinterwordspacing}{\spaceskip=0pt\relax}
\providecommand{\BIBentryALTinterwordstretchfactor}{4}
\providecommand{\BIBentryALTinterwordspacing}{\spaceskip=\fontdimen2\font plus
\BIBentryALTinterwordstretchfactor\fontdimen3\font minus
  \fontdimen4\font\relax}
\providecommand{\BIBforeignlanguage}[2]{{%
\expandafter\ifx\csname l@#1\endcsname\relax
\typeout{** WARNING: IEEEtran.bst: No hyphenation pattern has been}%
\typeout{** loaded for the language `#1'. Using the pattern for}%
\typeout{** the default language instead.}%
\else
\language=\csname l@#1\endcsname
\fi
#2}}
\providecommand{\BIBdecl}{\relax}
\BIBdecl

\bibitem{Vemp13}
A.~Vempaty, L.~Tong, and P.~Varshney, ``Distributed inference with byzantine
  data: State-of-the-art review on data falsification attacks,'' \emph{Signal
  Processing Magazine, IEEE}, vol.~30, no.~5, pp. 65--75, Sept 2013.

\bibitem{WSNvarshney}
R.~Niu and P.~K. Varshney, ``Distributed detection and fusion in a large
  wireless sensor network of random size,'' \emph{EURASIP Journal on Wireless
  Communications and Networking}, vol. 2005, no.~4, pp. 462--472, 2005.

\bibitem{WSNanomalyDet}
S.~Rajasegarar, C.~Leckie, M.~Palaniswami, and J.~Bezdek, ``Distributed anomaly
  detection in wireless sensor networks,'' in \emph{Proc. of 10th IEEE Intl.
  Conf. on Com. Systems}, Oct 2006, pp. 1--5.

\bibitem{WLSH10}
W.~Wang, H.~Li, Y.~Sun, and Z.~Han, ``Securing collaborative spectrum sensing
  against untrustworthy secondary users in cognitive radio networks,''
  \emph{EURASIP Journal on Advances in Signal Processing}, vol. 2010, p.~4,
  2010.

\bibitem{Raw11}
A.~S. Rawat, P.~Anand, H.~Chen, and P.~K. Varshney, ``Collaborative spectrum
  sensing in the presence of byzantine attacks in cognitive radio networks,''
  \emph{IEEE Transactions on Signal Processing}, vol.~59, no.~2, pp. 774--786,
  February 2011.

\bibitem{Mar09}
S.~Marano, V.~Matta, and L.~Tong, ``Distributed detection in the presence of
  byzantine attacks,'' \emph{IEEE Transactions on Signal Processing}, vol.~57,
  no.~1, pp. 16--29, Jan 2009.

\bibitem{Bar13}
M.~Barni and B.~Tondi, ``Multiple-observation hypothesis testing under
  adversarial conditions,'' in \emph{IEEE International Workshop on Information
  Forensics and Security (WIFS)}, Nov 2013, pp. 91--96.

\bibitem{OptFusion}
Z.~Chair and P.~Varshney, ``Optimal data fusion in multiple sensor detection
  systems,'' \emph{IEEE Transactions on Aerospace and Electronic Systems}, vol.
  AES-22, no.~1, pp. 98--101, Jan 1986.

\bibitem{Var97}
P.~K. Varshney, \emph{Distributed Detection and Data Fusion}.\hskip 1em plus
  0.5em minus 0.4em\relax Springer-Verlag, 1997.

\bibitem{CDC}
A.~Abrardo, M.~Barni, K.~Kallas, and B.~Tondi, ``Decision fusion with corrupted
  reports in multi-sensor networks: a game-theoretic approach,'' in \emph{IEEE
  Conference on Decision and Control (CDC), accepted paper}, Los Angeles,
  California, December 2014.

\bibitem{LearnByzantines}
A.~Vempaty, K.~Agrawal, P.~Varshney, and H.~Chen, ``Adaptive learning of
  byzantines' behavior in cooperative spectrum sensing,'' in \emph{Wireless
  Communications and Networking Conference (WCNC), 2011 IEEE}, March 2011, pp.
  1310--1315.

\bibitem{tolerant_scheme}
R.~Chen, J.-M. Park, and K.~Bian, ``Robust distributed spectrum sensing in
  cognitive radio networks,'' in \emph{INFOCOM 2008. The 27th Conference on
  Computer Communications. IEEE}, April 2008, pp.~--.

\bibitem{KBKV13}
B.~Kailkhura, S.~Brahma, Y.~Han, and P.~Varshney, ``Optimal distributed
  detection in the presence of byzantines,'' in \emph{ICASSP 2013, IEEE
  International Conference on Acoustics, Speech and Signal Processing}, May
  2013, pp. 2925--2929.

\bibitem{BarGon13}
M.~Barni and F.~{P{\'e}rez-Gonz{\'a}lez}, ``Coping with the enemy: advances in
  adversary-aware signal processing,'' in \emph{ICASSP 2013, IEEE International
  Conference on Acoustics, Speech and Signal Processing}, Vancouver, Canada,
  May 2013.

\end{thebibliography}

%
%
%
%
%
%
%
%

\end{document}